\documentclass{article}

\usepackage[margin=1.1in]{geometry}
\usepackage{amsmath}
\usepackage{amsfonts}
\usepackage{amssymb}
\usepackage{graphicx}
\usepackage{caption}
\usepackage{subcaption}
\usepackage{mathtools}
\usepackage{natbib}
\usepackage{hyperref}
\usepackage[capitalise]{cleveref}
\usepackage{fancyhdr}

\AddToHook{cmd/appendix/before}{\crefalias{section}{appendix}}

\graphicspath{{Figs/}}

\newcommand{\mathsfbi}[1]{\boldsymbol{\mathsf{#1}}}
\newcommand{\pder}[2]{\partial_{#2} #1}

\newcommand{\dint}{\,\textrm{d}}

\newcommand{\mat}[1]{\mathsfbi{#1}}

\renewcommand{\mat}[1]{\mathsfbi{#1}}

\let\P\relax
\DeclareMathOperator{\J}{J}
\DeclareMathOperator{\R}{R}
\DeclareMathOperator{\P}{P}

\counterwithin{equation}{section}

\title{\bf N-Layer quasi-geostrophic modons in a vertically sheared flow}
\author{Matthew N. Crowe\footnote{Matthew.Crowe2@ncl.ac.uk}$\hspace{5pt}^{1,2}$ \\
\small{$\,^{1}\,$School of Mathematics, Statistics and Physics, Newcastle University, Newcastle upon Tyne, NE1 7RU, UK} \\ \small{$\,^{2}\,$Department of Mathematics, University College London, London, WC1E 6BT, UK}}
\date{}
\date{}

\begin{document}

\maketitle

\begin{abstract}
Modons, or dipolar vortices, are a form of propagating vortex solution consisting of two counterrotating regions of fluid and moving together under self-advection. They are common features of the upper ocean that can remain stable over long times, thereby transporting fluid over large distances. Here, we present a semi-analytical method for finding modon solutions in a vertically-sheared background flow using the multi-layer quasi-geostrophic framework. This method works for arbitrarily many fluid layers and builds on our previous studies of multi-layer modon solutions. We present three example of modons in a background flow, highlighting the effect of varying relative potential velocity between layers and the impact of depth-dependent potential vorticity gradients arising from surface deformation and velocity shear.
\end{abstract}

\section{Introduction}

Modons, or dipolar vortices, are a form of coherent vortex structure consisting of two oppositely-signed regions of vorticity and propagating together through self-advection. They are common features of the ocean and typically remaining stable over long times, allowing for the transport of heat, nutrients and carbon over large distances \citep{NycanderI90, Hughes_Miller_2017}. Modons are typically observed at the ocean surface via satellite altimetry though they can extend over several kilometres into the deep ocean \citep{NiZWH20}. In the field of atmospheric science, modons have been used to model various phenomena---such as the Madden-Julien oscillation \citep{rostami_zeitlin_2021} and atmospheric blocking \citep{mcwilliams_1980}---while similar magnetohydrodynamical solutions may be relevant to stellar transport \citep{Lahaye_Zeitlin_2022}.

Analytical modon solutions were first independently discovered by Lamb and Chaplygin \citep{Lamb_1932,MeleshkoH94} as solutions to the two-dimensional Euler equations consisting of an isolated, circular dipolar core surrounded by a decaying far-field. These solutions were later generalised to the equivalent barotropic quasi-geostrophic equations on the beta-plane by \citet{LarichevR76d} and to the two-layer quasi-geostrophic equations by \citet{Kizner_et_al_2003}. In addition to two-dimensional modons, there has been much work on finding exact steady state vortex solutions across a range of fluid dynamical models; these solutions include the classical Hill's vortex \citep{HILL1894}, a broad class of spherical vortices discovered by \citet{MOFFATT69}, and the recent three-dimensional quasi-geostrophic vortices of \citet{Viudez_2019}.

Due to the difficulties with finding exact solutions to fully nonlinear equations, some studies have focussed on finding solutions using numerical methods. These methods include the iterative methods of \citet{Boyd_Ma_1990} for finding elliptical modons or the numerical transform based approach of \citet{MurakiS07} for finding modon solutions to the surface quasi-geostrophic model. Similarly, in cases where modon solutions cannot be found analytically, asymptotic theory has been used to demonstrate the existence of dipolar vortices and determine their properties \citep{fraenkel_1972,norbury_1972,norbury_1973}.

The stability properties of modons are not well understood and several early attempts at proving stability were shown to be incorrect by \citet{Nycander1992}. Subsequently, there has been much recent work on understanding modon breakdown processes; such as instabilities \citep{Davies_et_al_2023,Protas_2024,CroweSutyrin2025}, viscous dissipation \citep{FlorVD95,NielsenR97}, and energy loss due to wave radiation \citep{FlierlH94,JohnsonC21,Crowe_Johnson_23}. 

Recently, \citet{Johnson_Crowe_23} presented a new semi-analytical method for finding modon solutions in a surface quasi-geostrophic model assuming vanishing interior potential vorticity. This method used a Hankel transform and orthogonal polynomial expansion to determine exact solutions in terms of a set of unknown expansion coefficients. The expansion coefficients may be determined numerically by solving an inhomogeneous eigenvalue problem using standard methods. This method was later generalised to the multi-layer quasi-geostrophic system by \citet{Crowe_Johnson_2024} and the three-dimensional quasi-geostrophic system by \citet{Crowe_2026}. The multi-layer quasi-geostrophic equations may also be solved for modon solutions using the approach of \citet{Koval_et_al_2026}, who diagonalised the system and solved the resulting equations following \citet{LarichevR76d}. Exact analytical solutions were obtained in the case of a shared eigenbasis inside and outside the modon core.

Typically, modons are assumed to exist in the absence of a background flow. However, the ocean consists of many large-scale flow features---such as boundary currents and mesoscale eddies---which may alter the structure of propagating vortices. In this paper, we explore the effects of a vertically-sheared background flow on the structure of multi-layer quasi-geostrophic modons. We begin by generalising the multi-layer model and solution method of \citet{Crowe_Johnson_2024} to include a background flow. We also relax some of the assumptions made in that work by allowing a free surface in the upper layer and a non-constant density difference between layers. We then present three examples of modons in a background flow and discuss their structure and properties.

\section{Problem Setup}
\label{sec:setup}

Consider a cylindrical dipolar vortex of radius $a$ moving with speed $U$ in the zonal $(x)$ direction. We assume that the fluid region is rapidly rotating and may be divided into $N$-layers with approximately constant density in each layer. Assuming that the Rossby number of the flow is small, we make the quasi-geostrophic (QG) assumption that the flow velocity is primarily set by a balance between the Coriolis force and hydrostatic pressure gradients. In the frame of the moving vortex, we obtain the $N$-layer QG equations
\begin{equation}
\label{eq:QG_sys}
(\pder{}{t} - U \pder{}{x})q_i + J[\psi_i, q_i] + \beta_i \pder{\psi_i}{x} = 0, \quad i \in \{1, 2, \dots, N\},
\end{equation}
for streamfunction $\psi_i$ and potential vorticity anomaly $q_i$ in each layer. Here
\begin{equation}
J[\phi, \varphi] = \pder{\phi}{x}\,\pder{\varphi}{y} - \pder{\varphi}{x}\,\pder{\phi}{y}, 
\end{equation}
denotes the Jacobian derivative and $\beta_i$ denotes the background potential vorticity gradient in the meridional ($y$) direction. We allow $\beta_i$ to vary between layers to incorporate additional layer-dependent PV gradients, such as a bottom slope, in addition to the planetary beta effect. The potential vorticity anomaly in each layer may be determined from the streamfunction as
\begin{equation}
q_i = \nabla^2 \psi_i + \Lambda_{ij} \psi_j,
\end{equation}
where $\nabla^2 = \pder{^2}{x^2} + \pder{^2}{y^2}$ denotes the horizontal Laplacian and Einstein summation convention is used in the second term. The $N \times N$ tridiagonal matrix $[\mat{\Lambda}]_{ij} = \Lambda_{ij}$ represents the coupling between neighbouring layers through the deformation of the interface and is given by
\begin{equation}
\label{eq:R_def}
\mat{\Lambda} = \begin{pmatrix}
-\left[ R_0^{-2} + R_{3/2,1}^{-2}\right]\!\! & R_{3/2,1}^{-2} & 0 & \dots & 0 & 0 \\ R_{3/2,2}^{-2} & \!-\left[ R_{3/2,2}^{-2} + R_{5/2,2}^{-2} \right]\!\! & R_{5/2,2}^{-2} & \dots & 0 & 0 \\
0 & R_{5/2,3}^{-2} & \!-\left[ R_{5/2,3}^{-2} + R_{7/2,3}^{-2} \right]\!\! & \dots & 0 & 0 \\
\vdots & \vdots & \vdots & \ddots & \vdots & \vdots \\
0 & 0 & 0 & \dots & -R_{n-1/2,n}^{-2} & R_{n-1/2,n}^{-2}
\end{pmatrix},\!
\end{equation}
where
\begin{equation}
R_{i-1/2,j} = \sqrt{\frac{g(\rho_{i} - \rho_{i-1})H_j}{\rho_0 f^2}},
\end{equation}
denotes the internal (or Baroclinic) Rossby radius in layer $j$ resulting from the deformation of the interface between layers $i-1$ and $i$ and
\begin{equation}
R_0 = \sqrt{\frac{gH_1}{f^2}},
\end{equation}
denotes the barotropic Rossby radius of deformation describing deformations of the top free surface. Here $H_i$ and $\rho_i$ denote the depth and density of layer $i$ respectively, $g$ denotes the acceleration due to gravity, $\rho_0$ is the average fluid density, and $f$ is the Coriolis parameter describing the rotation of the system.

We now introduce a vertically sheared velocity in the $x$ direction by writing
\begin{equation}
\psi_i = \psi_i' - U_i y,
\end{equation}
where $U_i$ denotes a (constant) velocity in the $x$ direction in layer $i$ and $\psi_i'$ is a streamfunction describing the deviation in velocity from this background flow. Note that steady vortex solutions will not exist if the background flow has any component in the $y$ direction as the flow would generate a PV gradient in the direction of vortex propagation, resulting in a change of vortex structure with time to conserve PV. Substituting for $\psi_i$ in \cref{eq:QG_sys} gives
\begin{equation}
\label{eq:QG_sys_2}
(\pder{}{t} - U \pder{}{x})q_i' + J[\psi_i' - U_i y, q_i' - \Lambda_{ij} U_j y] + \beta_i \pder{\psi_i'}{x} = 0, \quad i \in \{1, 2, \dots, N\},
\end{equation}
where
\begin{equation}
q_i' = \nabla^2 \psi_i' + \Lambda_{ij}\psi_j'.
\end{equation}
\cref{eq:QG_sys_2} may be rewritten as
\begin{equation}
\pder{q_i'}{t} + J[\psi_i' + (U - U_i) y, q_i' + (\beta - \Lambda_{ij} U_j) y] = 0.
\end{equation}

We now look for steady, nonlinear modon solutions consisting of an isolated region of vorticity in the frame of the vortex for $x^2 + y^2 < a^2$. Taking $\pder{q_i'}{t} = 0$ gives that
\begin{equation}
\label{eq:Jac_psi_q}
J[\psi_i' + (U - U_i) y, q_i' + (\beta - \Lambda_{ij} U_j) y] = 0,
\end{equation}
and using the properties of the Jacobian derivative we have
\begin{equation}
\label{eq:functional_relation}
q_i' + (\beta - \Lambda_{ij} U_j) y = F_i\left( \psi_i' + (U - U_i) y \right),
\end{equation}
where $F_i$ is an arbitrary, piecewise differentiable function of a single variable. To proceed, we take $F_i$ to be a piecewise linear function
\begin{equation}
\label{eq:F_def}
F_i(\varphi) = \begin{cases}
-\frac{K_i^2}{a^2} \varphi, & x^2 + y^2 < a^2,\\
\frac{\beta_i - \Lambda_{ij}U_j}{U - U_i}\varphi, & x^2 + y^2 > a^2,
\end{cases}
\end{equation}
where $K_i$ is an unknown internal wavenumber describing the vortex core in each layer. The form of $F_i$ outside the vortex, $x^2 + y^2 > a^2$, is set by the requirement that the perturbation streamfunction, $\psi_i'$, and PV anomaly, $q_i'$, in each layer decay as $x^2+y^2 \to \infty$. We impose the condition
\begin{equation}
\label{eq:boundary_condition}
\Psi_i = \psi_i' + (U-U_i)y = 0 \quad\mathrm{on}\quad x^2 + y^2 = a^2,
\end{equation}
which ensures that the vortex boundary is a streamline in the frame of the vortex. Further, \cref{eq:boundary_condition} ensures that $F_i(\Psi_i)$ is continuous across the vortex boundary, allowing us to find modon solutions where the potential vorticity and streamfunction are also continuous across this boundary.

Before moving on to our solution method, it is useful to discuss the different streamfunctions and their interpretations. The streamfunction $\psi_i$ describes the flow in the laboratory frame and consists of a moving vortex and a background shear flow. The perturbation streamfunction $\psi_i'$ describes the vortex within each layer and consists of a localised region of vorticity moving with speed $U-U_i$ in the positive $x$ direction. Finally, the streamfunction $\Psi_i = \psi_i + Uy = \psi_i' + (U-U_i)y$ describes the flow in the frame of the moving vortex. It consists of a stationary vortex with vortex boundary given by the closed circular streamline $x^2 + y^2 = a^2$ and an oncoming background flow of speed $U - U_i$ in the negative $x$ direction.

In the case of vanishing background flow, $U_i = 0$ for $i \in \{1,2,\dots, N\}$, no surface deformation, $R_0 = \infty$, and constant density difference between layers, $\rho_i - \rho_{i-1} = \Delta\rho$ for $i \in \{2,3,\dots,N\}$, the model presented here reduces to the $N$-layer system solved in \citet{Crowe_Johnson_2024}.

\section{Solution Method}
\label{sec:method}

Following \citet{Crowe_Johnson_2024} we express the streamfunction in each layer using a Hankel transform
\begin{equation}
\label{eq:Hankel}
\psi_i' = U a \sin\theta \int_0^\infty \hat{\psi}_i(\xi) \J_1(\xi r / a) \xi \dint \xi,
\end{equation}
where $(r, \theta)$ are plane polar coordinates, $\J_1$ is the Bessel function of the first kind of order 1. Substituting \cref{eq:Hankel} into \cref{eq:functional_relation} and using the form of $F_i$ given in \cref{eq:F_def} gives
\begin{align}
\label{eq:interior_eqn}
\int_0^\infty \left[ \xi^2\mat{I} + \mat{\Gamma} - \mat{K}\right]\hat{\boldsymbol{\psi}}(\xi) \J_1(s\xi) \xi \dint \xi & = \left[ \mat{M} + \mat{K}\right] \boldsymbol{\gamma} s, & s < 1,\\
\label{eq:exterior_eqn}
\int_0^\infty \left[ \xi^2\mat{I} + \mat{\Gamma} + \mat{M} \right] \hat{\boldsymbol{\psi}}(\xi) \J_1(s\xi) \xi \dint \xi & = \textbf{0} & s > 1,
\end{align}
where $s = r/a$, $\mat{I}$ denotes the $N \times N$ identity matrix, and
\begin{equation}
\mat{\Gamma} = - a^2\mat{\Lambda}, \quad [\mat{K}]_{ij} = K_i^2 \delta_{ij},\quad [\mat{M}]_{ij} = \frac{(\beta_i - \Lambda_{ik}U_k)a^2}{U - U_i}\delta_{ij}, \quad [\hat{\boldsymbol{\psi}}(\xi)]_i = \hat{\psi}_i(\xi), \quad [\boldsymbol{\gamma}]_i = \frac{U - U_i}{U}.
\end{equation}
The continuity condition, $\Psi_i = 0$ on $s = 1$ from \cref{eq:boundary_condition}, can be written as
\begin{equation}
\label{eq:continuity}
\int_0^\infty \left[ \xi^2\mat{I} + \mat{\Gamma} + \mat{M} \right] \hat{\boldsymbol{\psi}}(\xi) \J_1(\xi) \xi \dint \xi = \textbf{0}.
\end{equation}
In addition to enforcing that the streamfunction is continuous, this condition also ensures that the vorticity is continuous, and hence the first two radial derivatives of $\psi_i'$ are continuous across the vortex boundary. This is consistent with previous studies of modon solutions in which the velocity and vorticity are continuous across the vortex boundary with a jump discontinuity appearing in the third radial derivative of the streamfunction \citep{LarichevR76d,MeleshkoH94}.

To proceed, we write
\begin{equation}
\label{eq:A_expansion}
\textbf{A}(\xi) = \left[ \xi^2\mat{I} + \mat{\Gamma} + \mat{M} \right] \hat{\boldsymbol{\psi}}(\xi) \xi = \sum_{j = 0}^\infty \textbf{a}_j \J_{2j + 2}(\xi),
\end{equation}
where the $\textbf{a}_j$ are vector-valued expansion coefficients. This choice automatically satisfies the governing equation in $s > 1$, \cref{{eq:exterior_eqn}}, using an integral relation given in \cref{eq:R_integral_1}. In the vortex interior, $s < 1$, we may substitute for $\hat{\boldsymbol{\psi}}$ using \cref{eq:A_expansion} in \cref{eq:interior_eqn} to obtain
\begin{equation}
\label{eq:inner_step_1}
\sum_{j = 0}^\infty \left[ \int_0^\infty \left[ \xi^2\mat{I} + \mat{\Gamma} - \mat{K} \right] \left[ \xi^2\mat{I} + \mat{\Gamma} + \mat{M} \right]^{-1} \J_{2j+2}(\xi) \J_1(s\xi) \dint \xi\right] \textbf{a}_j =  \left[ \mat{M} + \mat{K}\right] \boldsymbol{\gamma} s, \quad s > 1.
\end{equation}
We now introduce the \textit{Zernike radial functions} denoted by $\R_k(s)$, a set of orthogonal polynomials which have several useful properties described in \cref{app:Zernike}. Multiplying \cref{eq:inner_step_1} by $s\R_k(s)$, integrating over $s \in [0, 1]$, and using the Hankel transform result in \cref{eq:R_integral_2} gives
\begin{equation}
\label{eq:inner_step_2}
\sum_{j = 0}^\infty \left[ \int_0^\infty \left[ \xi^2\mat{I} + \mat{\Gamma} - \mat{K} \right] \left[ \xi^2\mat{I} + \mat{\Gamma} + \mat{M} \right]^{-1} \xi^{-1} \J_{2j+2}(\xi) \J_{2k+2}(\xi) \dint \xi\right] \textbf{a}_j =  \frac{\delta_{k0}}{4}\left[ \mat{M} + \mat{K}\right] \boldsymbol{\gamma},
\end{equation}
where $k \in \{0, 1, 2, 3 \dots\}$. Defining
\begin{align}
\label{eq:A_kj}
\mat{A}_{kj} & = \int_0^\infty \left[ \xi^2\mat{I} + \mat{\Gamma}\right] \left[ \xi^2\mat{I} + \mat{\Gamma} + \mat{M} \right]^{-1} \xi^{-1} \J_{2j+2}(\xi) \J_{2k+2}(\xi) \dint \xi,\\
\label{eq:B_kj}
\mat{B}_{kj} & = \int_0^\infty \left[ \xi^2\mat{I} + \mat{\Gamma} + \mat{M} \right]^{-1} \xi^{-1} \J_{2j+2}(\xi) \J_{2k+2}(\xi) \dint \xi,\\
c_k & = \frac{\delta{k0}}{4},
\end{align}
allows us to write \cref{eq:inner_step_2} in the form of an inhomogeneous eigenvalue problem as
\begin{equation}
\label{eq:EVP}
\sum_{j = 0}^\infty \left[ \mat{A}_{kj} - \mat{K} \mat{B}_{kj}\right] \textbf{a}_j = \left[ \mat{M} + \mat{K}\right] \boldsymbol{\gamma} c_k.
\end{equation}
The continuity condition, \cref{eq:continuity}, gives that
\begin{equation}
\int_0^\infty \textbf{A}(\xi) \J_1(\xi) \dint \xi = \sum_{j = 0}^\infty \left[ \int_0^\infty \J_{2j+2}(\xi) \J_1(\xi) \dint \xi \right]\textbf{a}_j = \sum_{j = 0}^\infty \R_j(1) \textbf{a}_j = \sum_{j = 0}^\infty (-1)^j \textbf{a}_j = \textbf{0},
\end{equation}
using the result $\R_n(1) = (-1)^n$. This expression gives an extra equation linking the coefficients, $\textbf{a}_j$, and is required to find unique solutions to \cref{eq:EVP}.

The integrals $\mat{A}_{kl}$ and $\mat{B}_{kj}$ may be evaluated numerically by deforming the integration contour into the complex plane such that any oscillatory components decay exponentially \citep{Crowe_Johnson_23}. Should the function $[\xi^2\mat{I} + \mat{\Gamma} + \mat{M}]^{-1}$ have any singularities along the positive real axis, the values of $\mat{A}_{kl}$ and $\mat{B}_{kj}$ cannot be uniquely determined. This non-uniqueness corresponds to the existence of stationary Rossby wave modes with a phase speed that matches the propagation speed of the vortex. In this case, steady vortex solutions will not exist as the Rossby wave-field will extract energy from the vortex resulting in a change of vortex speed and/or structure over time \citep{FlierlH94,JohnsonC21}. Typically, Eastward ($U > 0$) modon solutions will exist while Westward ($U < 0$) modons will only exist if they travel faster than all Rossby wave modes.

\cref{eq:EVP} may be truncated after $M$ terms by taking $j, k \in \{0, 1, 2, \dots, M-1\}$, and solved for eigenvalues, $K_i^2$, and coefficient vectors, $\textbf{a}_j$, using the approach described in \citet{Crowe_Johnson_2024}. Once the coefficients are determined, we can calculate the vortex streamfunction, $\boldsymbol{\psi}$, and potential vorticity anomaly, $\boldsymbol{q}$, defined as $[\boldsymbol{\psi}]_i = \psi_i'$ and $[\boldsymbol{q}]_i = q_i$. Starting from the Hankel transform of $\boldsymbol{\psi}$ given in \cref{eq:Hankel}, we can write
\begin{equation}
\left[ \nabla^2 \mat{I} + \mat{\Lambda} - \frac{1}{a^2}\mat{M} \right] \boldsymbol{\psi} = - \frac{U}{a} \sin\theta \int_0^\infty \left[ \xi^2 +\mat{\Gamma} + \mat{M}\right] \hat{\boldsymbol{\psi}}(\xi) \J_1(s\xi) \xi \dint \xi,
\end{equation}
and using \cref{eq:A_expansion,eq:R_integral_1} we can show that
\begin{equation}
\label{eq:psi_sol}
\left[ \nabla^2 \mat{I} + \mat{\Lambda} - \frac{1}{a^2}\mat{M} \right] \boldsymbol{\psi} = - \frac{U}{a} \sin\theta \sum_{j = 0}^\infty \textbf{a}_j \R_j(r/a)\Theta(a - r),
\end{equation}
where $\Theta(z)$ denotes the Heaviside step function. Typically, the quickest way to determine $\boldsymbol{\psi}$ is to invert \cref{eq:psi_sol} numerically at each point in Fourier space using discrete Fourier transforms over a given $(x, y)$ grid. The potential vorticity anomaly may then be calculated using
\begin{equation}
\label{eq:q_sol}
\boldsymbol{q} = \left[ \nabla^2 \mat{I} + \mat{\Lambda} \right] \boldsymbol{\psi},
\end{equation}
which may be similarly evaluated in Fourier space.

\section{Examples}
\label{sec:examples}

We now present three examples that demonstrate how the inclusion of a background shear flow can modify the structure of three-dimensional modons. Here, we will consider only Eastward propagating modons as Westward steady-state solutions may not exist due to the generation of a Rossby wave wake \citep{FlierlH94, JohnsonC21} that extracts energy from the vortex.

\subsection{Example 1: Constant flow and a finite barotropic Rossby radius}

We begin by considering a three-layer beta-plane modon in a constant background flow. A bottom slope is included in the lowest layer such that the value of $\beta_3$ is different from $\beta_i$ for $i = 1,2,3$. We take $(U, a) = (1, 1)$, $(\beta_1, \beta_2, \beta_3) = (1, 1, 0)$, and $(U_1, U_2, U_3) = (0.5, 0.5, 0.5)$ and include a barotropic Rossby radius, $\R_0 = 1$, in the top layer and baroclinic Rossby radii, $(R_{3/2,1}, R_{3/2,2}, R_{5/2,2}, R_{5/3,3}) = (1, 1, 1, 1)$.

\begin{figure}[t!]
\centering
\includegraphics[width=\linewidth]{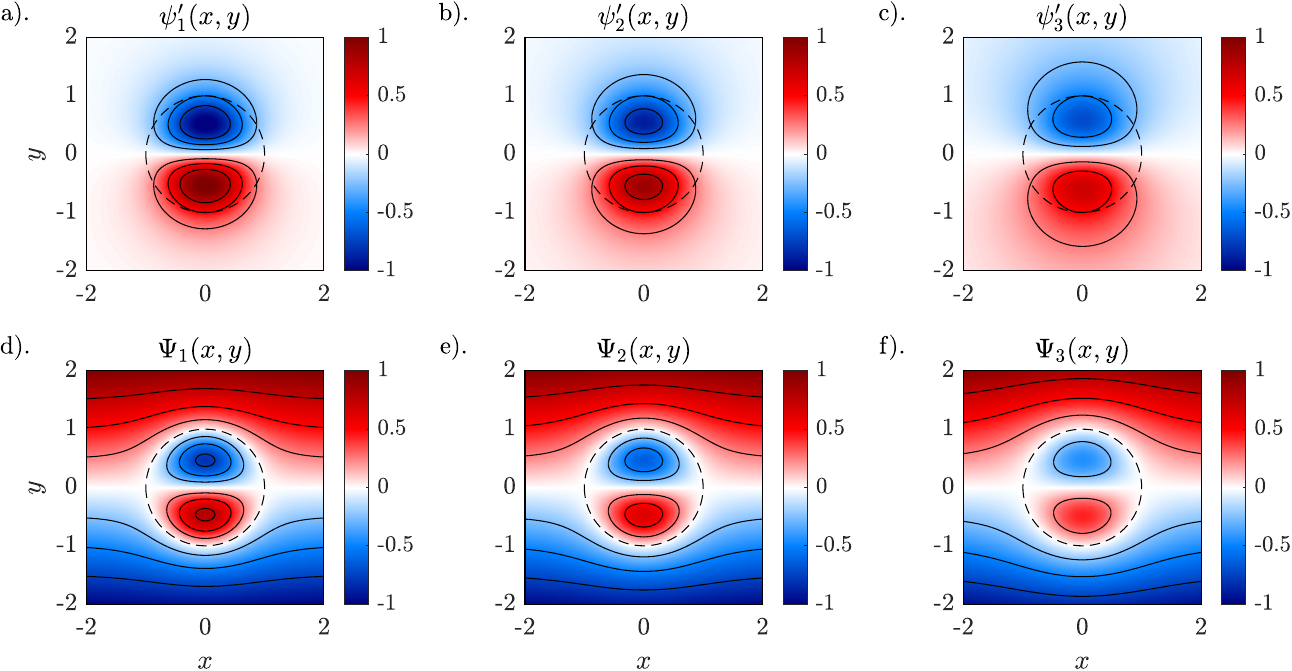}
\caption{Plots of the perturbation streamfunction, $\psi_i'$, (panels a-c) and the streamfunction in the frame of the vortex, $\Psi_i$, (panels d-f) in each layer as functions of horizontal position $(x, y)$. The vortex core boundary, $r = a$, is shown as a dashed circle and the solid contours denote level curves of the streamfunction (streamlines) at $(\pm 0.25, \pm 0.5 \pm 0.5)$. Results are shown for $N = 3$, $(U, a) = (1,1)$, $\beta_i = (1, 1, 0)$, $U_i = 0.5$, $R_0 = 1$ and $R_{i-1/2,j} = 1$.}
\label{fig:Ex_1}
\end{figure}

For these parameters, the background PV gradient is given by
\begin{equation}
\beta_i' = \beta_i - \Lambda_{ij}U_j = [(3/2, 1, 0)]_i.
\end{equation}
Since the background flow is independent of depth, the only non-zero component of $\Lambda_{ij}U_j$ arises from the barotropic Rossby radius term in the top layer. As $R_0$ is finite, the top surface can deform, resulting in a surface height gradient forming to balance the background flow through geostrophic balance. This height gradient generates a PV gradient via vortex stretching, thus modifying the background PV in the top layer only. We conclude that the simple addition of a free surface and bottom slope allows the PV gradient to differ across all three layers without needing a vertical shear in the background flow. We might expect modon solutions of speed $U = 1$ in a constant background flow of speed $U_i = 0.5$ to be identical to modons of speed $U = 0.5$ with no background flow due to Galilean invariance. However, as noted above, a background flow in the top layer drives a surface height gradient which breaks the invariance. As such, solutions in this example will be equivalent to modons with no background flow \citep{Crowe_Johnson_2024} only if the additional PV gradient is taken into account, i.e. by using $\beta'_i$ instead of $\beta_i$.

\cref{fig:Ex_1} shows the perturbation streamfunction, $\psi_i'$, and the streamfunction in the frame of the vortex, $\Psi_i$, in each layer. We can see that $\Psi_i = 0$ on the vortex boundary, $r = a$, consistent with the requirement that the vortex boundary is a streamline. Further, we observe that streamfunction contours are closest together in the top layer ($i = 1$), corresponding to higher velocities. This occurs due to vortex stretching; the pressure within the top layer creates a changes in surface height that is balanced by an increased geostrophic velocity compared to the other layers. The rate of decay of the perturbation streamfunction into the far field is also largest in the top layer and smallest in the bottom layer. From \cref{eq:functional_relation,eq:F_def} we can estimate this decay rate as $\lambda_i = \sqrt{-\Lambda_{ii} + \beta_i'}$ where the diagonal entries of $\Lambda_{ij}$ are negative. Due to the barotropic Rossby radius term in the top layer and the higher PV gradient in the upper layers, we expect the decay rate to decrease with depth, as shown in \cref{fig:Ex_1}.

\subsection{Example 2: A prograde linear shear flow}

\begin{figure}[t!]
\centering
\includegraphics[width=0.85\linewidth]{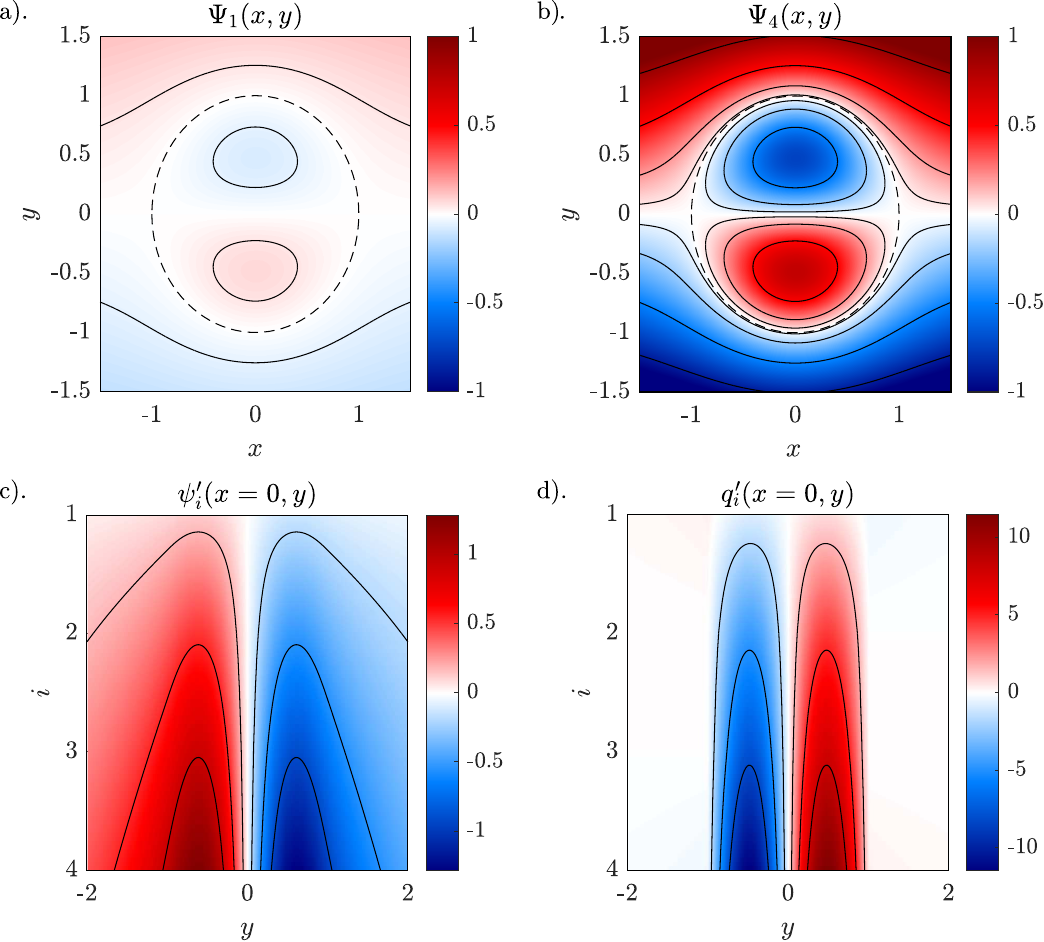}
\caption{Panels (a) and (b): the streamfunction in the frame of the vortex, $\Psi_i$, as a function of $(x,y)$ in the top (a) and bottom (b) layers. Panels (c) and (d): the perturbation streamfunction $\psi_i$ (c) and perturbation potential vorticity $q_i'$ (d) along the line $x = 0$ as functions of $y$ and layer number $i$. Results are shown for $N = 4$, $(U, a) = (1,1)$, $\beta_i = 0$, $U_i = (0.9, 0.6, 0.3, 0)$, $R_0 = \infty$ and $R_j = (1, 2, 2, 1)$.}
\label{fig:Ex_2}
\end{figure}

We now consider a four-layer f-plane modon moving in the same direction as a linearly sheared background flow. We take $(U, a) = (1, 1)$, $\beta_i = 0$, and $U_i = [(0.9, 0.6, 0.3, 0)]_i$ and assume no surface deformation, $R_0 = \infty$. We assume that the density change across layer interfaces is the same for each interface, $R_j = R_{j-1/2,j} = R_{j+1/2,j}$, and take $R_j = [(1, 2, 2, 1)]_j$. For this choice of baroclinic Rossby radii and background flow, the background PV gradient is given by
\begin{equation}
\beta_i' = - \Lambda_{ij}U_j = [(0.3, 0, 0, -0.3)]_i,
\end{equation}
hence a linear shear flow only generates a PV gradient in the top and bottom layers. This occurs as $\Lambda_{ij}$ approximates a second order vertical derivative in the interior layers for $R_{j-1/2,j} = R_{j+1/2,j}$, similarly to the 3D QG system \citep{Crowe_2026}.

The speed of the vortex in each layer relative to the background flow is $U - U_i = [(0.1, 0.4, 0.7, 1)]_i$ so we expect the magnitude of the perturbation streamfunction to increase with layer number as lower layers require stronger self-advection for the vortex to move at speed $U - U_i$. \cref{fig:Ex_2}.(a) and \cref{fig:Ex_2}.(b) show the streamfunction, $\Psi_i$, in the top and bottom layers respectively. We observe that, as expected, the streamfunction magnitude is significantly smaller in the top layer than the lower layer and the component describing the relative flow, $(U - U_i)y$, is weaker. \cref{fig:Ex_2}.(c) and \cref{fig:Ex_2}.(d) show the perturbation streamfunction, $\psi'_i$, and perturbation PV, $q_i'$ along $x = 0$ as functions of $y$ and layer number $i$. We again observe that the vortex strength increases with depth. We also note that there is a weak PV signature outside the vortex core, $r > a$, only in the top and bottom layers. This is consistent with the vanishing background PV gradient in the interior layers as fluid elements deflected around the vortex will only change their PV when moving up or down a background PV gradient.

\subsection{Example 3: A retrograde quadratic shear flow}

\begin{figure}[t!]
\centering
\includegraphics[width=0.85\linewidth]{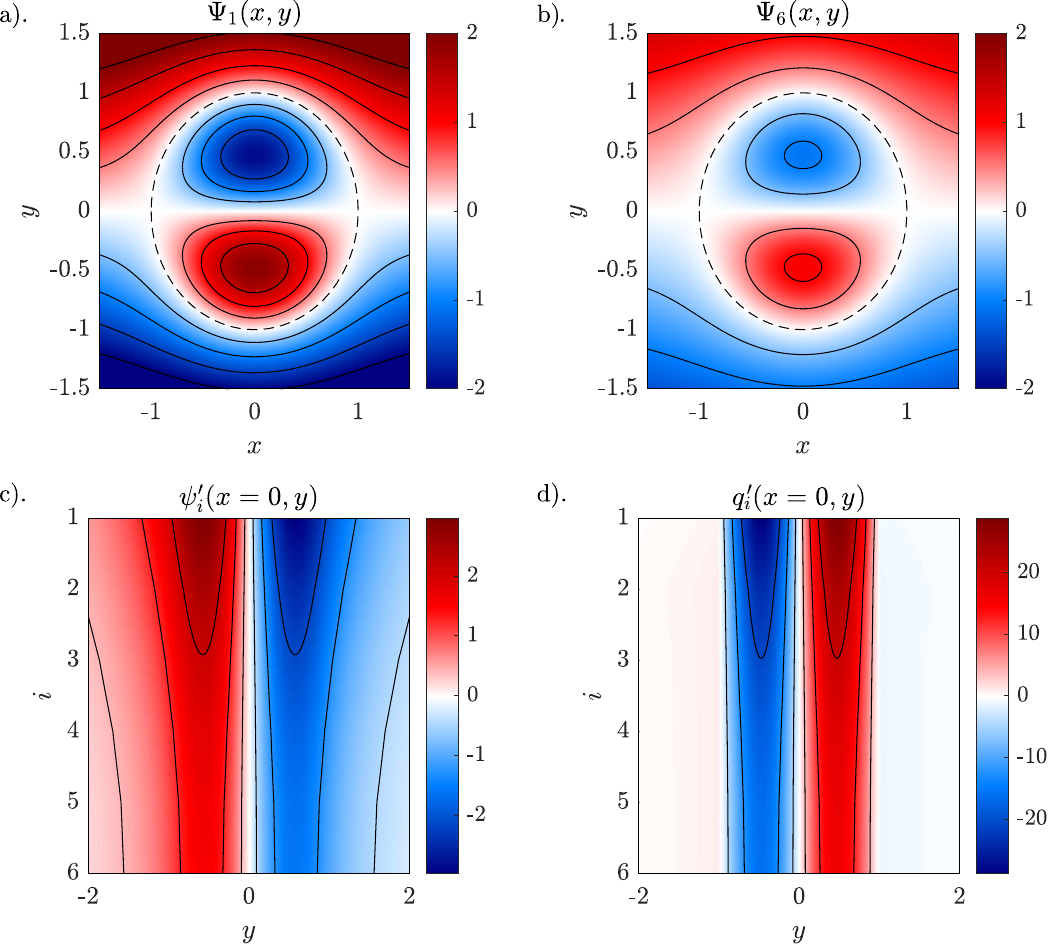}
\caption{Panels (a) and (b): the streamfunction in the frame of the vortex, $\Psi_i$, as a function of $(x,y)$ in the top (a) and bottom (b) layers. Panels (c) and (d): the perturbation streamfunction $\psi_i$ (c) and perturbation potential vorticity $q_i'$ (d) along the line $x = 0$ as functions of $y$ and layer number $i$. Results are shown for $N = 6$, $(U, a) = (1,1)$, $\beta_i = 1$, $U_i = -(1, 0.64, 0.36, 0.16, 0.04, 0)$, $R_0 = \infty$ and $(R_{3/2,1}, R_{3/2,2}, R_{5/2,2}, R_{5/2,3}, R_{7/2,3}, R_{7/3,4}, R_{9/2,4}, R_{9/2,5}, R_{11/2,5}, R_{11/2,6}) = (1, 2, 3, 3, 4, 3, 4, 2, 3, 1)$.}
\label{fig:Ex_3}
\end{figure}

Finally, we consider a six-layer beta-plane modon moving against a quadratically sheared background flow. We take $(U, a) = 1$, $\beta_i = 1$ and $U_i = -(6 - i)^2/25 = -[(1, 0.64, 0.36, 0.16, 0.04, 0)]_i$, and assume no surface deformations, $R_0 = \infty$. We assume that the density difference across interfaces varies between layers, $R_{j-1/2,j} \neq R_{j+1/2,j}$, and take
\begin{equation}
\begin{pmatrix}
R_{3/2,1} & R_{5/2,2} & R_{7/2,3} & R_{9/2,4} & R_{11/2,5} \\
R_{3/2,2} & R_{5/2,3} & R_{7/3,4} & R_{9/2,5} & R_{11/2,6}
\end{pmatrix} = \begin{pmatrix}
1 & 3 & 4 & 4 & 3 \\
2 & 3 & 3 & 2 & 1
\end{pmatrix}.
\end{equation}
This choice of baroclinic Rossby radius and background flow leads to different, non-zero, background PV gradients in each layer, with variations arising due to varying velocity shear across layer interfaces.

Since the background flow is surface-intensified and moving against the direction of propagation of the modon, we expect the streamfunction to have the largest magnitude in the upper layers as stronger self advection is required for the vortex to move through the oncoming flow. \cref{fig:Ex_3}.(a) and \cref{fig:Ex_3}.(b) show the streamfunction, $\Psi_i$, in the top and bottom layers respectively. As expected, we observe that the streamfunction magnitude is stronger in the top layer than the bottom layer and the component representing the background flow relative to the vortex, $(U - U_i)y$, is larger. \cref{fig:Ex_2}.(c) and \cref{fig:Ex_2}.(d) show the perturbation streamfunction, $\psi'_i$, and perturbation PV, $q_i'$ along $x = 0$ as functions of $y$ and layer number $i$.  We observe that the perturbation streamfunction magnitude decreases approximately quadratically with depth---consistent with the form of $(U - U_i)$---and the perturbation PV has a non-zero component outside the vortex $(r > a)$ describing how fluid parcels outside the vortex change PV as they move up/down the background PV gradient in each layer when deflected around the vortex core.

\section{Discussion and Conclusions}
\label{sec:conc}

We have presented a semi-analytical method for obtaining modon, or dipolar vortex, solutions to the multi-layer quasi-geostrophic model in the presence of a background shear flow. Such steady solutions will only exist if any potential vorticity (PV) gradients induced by the background flow, planetary rotation, or topographic slope are perpendicular to the direction of vortex propagation. This method generalises the results of \citet{Crowe_Johnson_2024} to include background flow, surface deformation, and varying density differences between layers. The variation of relative modon velocity between layers leads to a strong depth-dependence in the modon streamfunction and vorticity fields, while the background PV changes in each layer induced by the velocity shear across layer interfaces modifies both the vortex decay scale per layer and the PV change of fluid parcels deflected around the vortex core.

The existence of these steady modon solutions requires that there are no Rossby waves with phase speeds matching the propagation speed of the vortex. Typically, this means that Eastward propagating modons will always exist and Westward modons will only exist if travelling faster than the fastest Rossby wave mode, however the inclusion of background velocity shear and a topographic slope may lead to Eastward wave propagation under some parameter regimes. The appearance of singularities in the integrals given in \cref{eq:A_kj,eq:B_kj} indicates that a steady solution will not exist due to the appearance of these Rossby wave modes and can be used as a non-existence condition.

Vortex solutions obtained using our method may be used for theoretical and numerical studies of quasi-geostrophic vortex dynamics, or to create geostrophically balanced initial conditions for strongly rotating primitive equation models. While the solutions presented here are steady solutions to the multi-layer quasi-geostrophic system, we have not explored the stability of these solutions. Work by several authors suggests that modons may be susceptible to a range of destabilising processes, such as trajectory instabilities \citep{NycanderI90,Nycander1992} and symmetry breaking instabilities \citep{Davies_et_al_2023,CroweSutyrin2025}. As such, further study is required to determine the long-term evolution of the solutions presented here. Even steady modons are unlikely to remain steady once ageostrophic effects, dissipation, waves, and parametrisation terms are taken into account. Therefore, the examination of quasi-geostrophic modons in a realistic ocean or atmosphere model remains an important avenue for future research.

Finally, it should be noted that the multi-layer quasi-geostrophic system is susceptible to baroclinic instability in the presence of a vertically-sheared background flow \citep{Phillips_1951, Phillips_1954}. Therefore, propagating modons may be destabilised by growing baroclinic waves, or undergo complicated interactions with baroclinic eddies.
\\\\
\noindent{\bf Code Availability\bf{.}} MATLAB scripts implementing the method described in this paper and demonstrating the examples presented in \cref{sec:examples} are available \href{https://github.com/mncrowe/QGDipoles_v2.m}{here}.
\\\\
\noindent{\bf Declaration of Interests\bf{.}} The author reports no conflict of interest.
\\\\
\noindent{\bf AI Statement\bf{.}} No artificial intelligence tools were used in the preparation of this manuscript.

\appendix

\section{Zernike polynomials and integral relations}
\label{app:Zernike}

The \textit{Zernike radial functions}, $\R_n(s)$, are a set of degree $2n+1$ polynomials which are orthogonal over the interval $s \in [0, 1]$ with weight function $s$ \citep{BornW19}. The orthogonality relation is
\begin{equation}
\int_0^1 \R_n(s)\R_m(s) s \dint s = \frac{1}{4(n+1)}\,\delta_{nm}.
\end{equation}
Zernike radial functions may be expressed in terms of Jacobi polynomials, $\P_n^(\alpha,\beta)$, as
\begin{equation}
\R_n(s) = (-1)^n s\,\P_n^{(0,1)}(2s^2 - 1).
\end{equation}
A useful special case of this expression is $\R_n(1) = (-1)^n$. The Hankel transform of the function $\J_{2n+2}(\xi)/\xi$ allows us to relate integrals of Bessel functions to the Zernike radial function via
\begin{equation}
\label{eq:R_integral_1}
\int_0^\infty \J_{2n+2}(\xi)\J_1(s\xi) \dint \xi = \begin{cases}
\R_n(s) & s \leq 1,\\
0 & s > 1,\\
\end{cases}
\end{equation}
and the corresponding inverse Hankel transform gives
\begin{equation}
\label{eq:R_integral_2}
\J_{2n+2}(\xi)/\xi = \int_0^1 s \R_n(s) \J_1(s\xi) \dint s.
\end{equation}

\bibliographystyle{jfm}
\bibliography{bibliography}

\end{document}